\documentclass[journal]{IEEEtran}

\usepackage{amssymb}
\usepackage{amsmath}
\usepackage{cite}
\usepackage{url}
\usepackage{empheq}
\usepackage{xcolor}
\usepackage{graphicx}
\usepackage{subfigure}
\usepackage{enumitem}
\usepackage{fancyhdr}
\usepackage{mdwmath}	
\usepackage{mdwtab}
\usepackage{caption}
\usepackage{amsthm}
\usepackage{algorithm}
\usepackage{algorithmic}

\newtheorem{lemma}{Lemma}

\newtheorem{theorem}{Theorem}
\newtheorem{corollary}{Corollary}
\newtheorem{assumption}{Assumption}
\newtheorem{definition}{Definition}

\newcommand{\eqr}[1]{(\ref{#1})}
\newcommand{\fref}[1]{Fig.~\ref{#1}}

\begin{document}
\title{Antenna Activation for NOMA Assisted Pinching-Antenna Systems}
\author{Kaidi~Wang,~\IEEEmembership{Member,~IEEE,}
Zhiguo~Ding,~\IEEEmembership{Fellow,~IEEE,}
and Robert~Schober,~\IEEEmembership{Fellow,~IEEE}
\thanks{K. Wang and Z. Ding are with the Department of Electrical and Electronic Engineering, the University of Manchester, M1 9BB Manchester, UK (email: kaidi.wang@ieee.org; zhiguo.ding@ieee.org).}
\thanks{Z. Ding is also with Khalifa University, Abu Dhabi, UAE.}
\thanks{R. Schober is with the Institute for Digital Communications, Friedrich-Alexander-University Erlangen-Nurnberg (FAU), 91054 Erlangen, Germany  (e-mail: robert.schober@fau.de).}}
\maketitle
\setlength{\abovedisplayskip}{2pt}
\setlength{\belowdisplayskip}{2pt}
\begin{abstract}
In this letter, a non-orthogonal multiple access (NOMA) assisted downlink pinching-antenna system is investigated, where multiple pinching antennas can be activated at pre-configured positions along a dielectric waveguide to serve users via NOMA. In particular, the objective of this letter is to study at what locations and how many pinching antennas should be activated in order to maximize the system throughput. To this end, a sum rate maximization problem with antenna activation is formulated. With the help of matching theory, the formulated problem can be recast as a one-sided one-to-one matching, for which a low-complexity algorithm is developed. Simulation results indicate that the considered NOMA assisted pinching-antenna system can outperform conventional fixed-antenna systems in terms of sum rate, and the proposed matching based antenna activation algorithm yields a significant performance gain over the considered benchmarks.
\end{abstract}
\vspace{-2mm}
\section{Introduction}
Recently, flexible-antenna systems, including intelligent reflecting surfaces, fluid-antenna systems, and movable antennas, have received considerable attention \cite{wu2021intelligent, wong2020fluid, zhu2023modeling}. Due to their characteristics of dynamically reconfiguring wireless channels, flexible-antenna systems can yield significant performance gains over conventional fixed-location antenna systems. However, in most existing flexible antenna systems, the variation range of the antenna positions is limited to the wavelength scale, which restricts their capabilities to combat large-scale path loss \cite{ding2024pin}. Furthermore, the prohibitive cost of many existing flexible-antenna systems also limits their applications in real-world scenarios. In this context, pinching-antenna systems have been proposed, where low-cost dielectric materials, such as plastic clothespins, can be applied at any location on dielectric waveguides \cite{suzuki2022pinching} to create new line-of-sight links and/or enhancing existing transceiver channels. As a result, pinching antennas are expected to bring revolutionary breakthroughs in wireless communications.

\begin{figure}[!t]
\centering{\includegraphics[width=75mm]{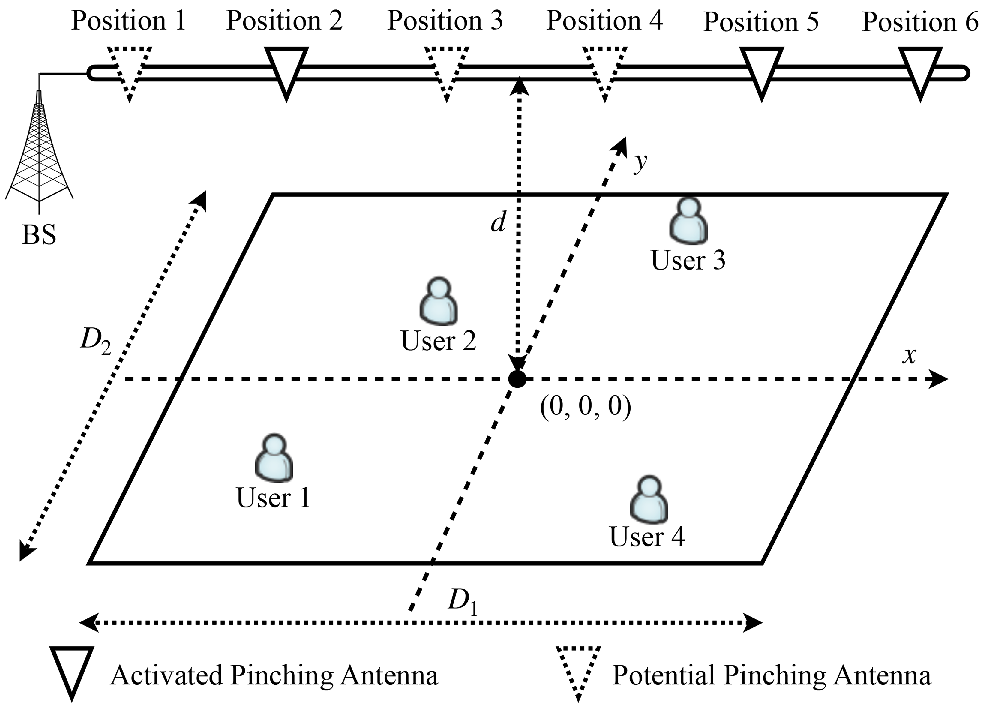}}
\caption{An illustration of the considered NOMA assisted pinching-antenna system with $N=4$ users and $L=6$ available antenna positions. The set of positions occupied by activated antennas is $\mathcal{S}=\{2,5,6\}$.}
\label{system}
\vspace{-4mm}
\end{figure}

One feature of pinching-antenna systems is that the pinching antennas activated on a given waveguide must transmit the same signal, which motivates the application of non-orthogonal multiple access (NOMA) \cite{ding2024pin}. In \cite{ding2024pin}, a NOMA assisted downlink pinching-antenna system was designed, where a heuristic approach was proposed to activate multiple pinching antennas along one waveguide in order to serve users simultaneously. As pointed out in \cite{ding2024pin}, the optimal performance of pinching-antenna systems can be achieved if these antennas can be activated at any desired position, which is challenging to realize in practice. Motivated by this, this letter focuses on a low-complexity and practical implementation, which is to install pinching antennas at preconfigured positions prior to transmission, and then activate only a small number of the available antennas to serve downlink users, as shown in \fref{system}. In particular, by transforming the antenna positions from continuous variables to discrete variables, a sum rate maximization problem is formulated and a matching based antenna activation algorithm is developed. The properties of the proposed algorithm, including its computational complexity, convergence, and stability, are analyzed. Simulation results demonstrate that i) the considered NOMA assisted pinching-antenna system is able to achieve higher spectral efficiency compared to conventional fixed-antenna systems, and ii) the proposed antenna activation algorithm can significantly improve the sum rate with low complexity, achieving near-optimal performance.

\vspace{-2mm}
\section{System Model and Problem Formulation}
Consider a downlink pinching-antenna system, where one base station (BS) serves $N$ single-antenna users with one waveguide and $K$ pinching antennas. As shown in \fref{system}, the users are randomly distributed within a rectangular area with side lengths $D_1$ and $D_2$, and the pinching antennas can be activated at $L$ antenna positions on a waveguide of length $D_1$ and deployed at height $d$, where $K\le L$. Note that the antenna spacing is jointly determined by the number of antenna positions and the length of the waveguide. Specifically, the antenna positions (or potential antennas) are uniformly spaced along the waveguide, and the distance between any two adjacent positions is $D_1/(L-1)$, which should be greater than or equal to half of the wavelength. In this letter, we make the practical assumption that the $K$ pinching antennas must be located at the pre-configured $L$ positions, and the antennas can be deactivated\footnote{One practical implementation of pinching antennas is to first build a track parallel to the waveguide, and install separate dielectrics, i.e., pinches, at the $L$ preconfigured positions. Similar to keys of self-playing pianos, the $K$ antennas can be activated by applying these pinches to the waveguide.}. To facilitate system description and algorithm design, the collections of all pinching antennas, users, and antenna positions are represented by $\mathcal{K}=\{1,2,\cdots,K\}$, $\mathcal{N}=\{1,2,\cdots, N\}$, and $\mathcal{L}=\{1,2,\cdots, L\}$, respectively.
\vspace{-2mm}
\subsection{NOMA Assisted Pinching-Antenna System}
In the considered scenario, the locations of user $n$ and the pinching antenna activated at position $l$ are denoted by $\psi_n=(x_n,y_n,0)$ and $\psi_l^\mathrm{Pin}=(x_l^\mathrm{Pin},0,d)$, respectively. With $L$ potential antennas, by using the spherical wave channel model, user $n$'s channel vector can be expressed as follows \cite{zhang2022nf}:
\begin{equation}
\mathbf{h}_n = \!\!\left[\frac{\eta e^{-j\frac{2\pi}{\lambda}\left|\boldsymbol{\psi}_n-\boldsymbol{\psi}_1^\mathrm{Pin}\right|}}{\left|\boldsymbol{\psi}_n-\boldsymbol{\psi}_1^\mathrm{Pin}\right|} \quad \cdots \quad \frac{\eta e^{-j\frac{2\pi}{\lambda}\left|\boldsymbol{\psi}_n-\boldsymbol{\psi}_L^\mathrm{Pin}\right|}}{\left|\boldsymbol{\psi}_n-\boldsymbol{\psi}_L^\mathrm{Pin}\right|}\right]^T,
\end{equation}
where $\eta=\frac{c}{4\pi f_c}$, $c$ is the speed of light, $f_c$ is the carrier frequency, and $\lambda$ is the wavelength. Since all pinching antennas are located on the same waveguide, there is a single radio frequency chain, and hence, the signals to be sent to the users have to be superimposed first before their transmission, which motivates the use of NOMA. That is, the superimposed signal can be expressed as $s=\sum_{n=1}^N\sqrt{\alpha_n}s_n$, where $\alpha_n$ and $s_n$ are the power allocation coefficient and the desired signal of user $n$, respectively. By taking into consideration the phase shifts caused by the signal propagation inside the waveguide \cite{ding2024pin}, the transmitted signal vector can be expressed as follows:
\begin{equation}
\mathbf{x}=\left[\sqrt{p_1}e^{-j\theta_1} \quad \cdots \quad \sqrt{p_L}e^{-j\theta_L}\right]^Ts,
\end{equation}
where $p_l$ is the transmit power of the antenna activated at position $l$, $\theta_l=2\pi\frac{\left|\boldsymbol{\psi}_0^\mathrm{Pin}-\boldsymbol{\psi}_l^\mathrm{Pin}\right|}{\lambda_g}$ is the corresponding phase shift due to the signal traveling within the waveguide, $\boldsymbol{\psi}_0^\mathrm{Pin}$ is the location of the feed point of the waveguide, $\lambda_g=\lambda/n_\mathrm{eff}$ is the guided wavelength, and $n_\mathrm{eff}$ is the effective refractive index of the dielectric waveguide \cite{pozar2021microwave}. At user $n$, the received signal is given by $y_n=\mathbf{h}_n^H\mathbf{x}+w_n$, which can be rewritten as follows:
\begin{equation}
y_n=\left(\sum_{l\in\mathcal{L}}\frac{\eta e^{-j\frac{2\pi}{\lambda}\left|\boldsymbol{\psi}_n-\boldsymbol{\psi}_l^\mathrm{Pin}\right|}}{\left|\boldsymbol{\psi}_n-\boldsymbol{\psi}_l^\mathrm{Pin}\right|}e^{-j\theta_l}\sqrt{p_l}\!\right)\!s+w_n,
\end{equation}
where $w_n$ is the additive white Gaussian noise.
\vspace{-2mm}
\subsection{Antenna Activation Principle and SIC Decoding Design}
In this paper, we consider the case where a subset of pinching antennas are activated from the $L$ available positions, and the set of all activated pinching antennas' positions is denoted by $\mathcal{S}$. In this case, $\mathcal{S}$ becomes a subset of all available positions, i.e., $\mathcal{S}\subseteq\mathcal{L}$, and the cardinality of $\mathcal{S}$ is less than or equal to the number of all antennas, i.e., $|\mathcal{S}|\le K$. By considering antenna activation, the received signal at user $n$ can be expressed as follows:
\begin{align}\nonumber
y_n(\mathcal{S})&=\left(\sum_{l\in\mathcal{S}}\frac{\eta e^{-j\frac{2\pi}{\lambda}\left|\boldsymbol{\psi}_n-\boldsymbol{\psi}_l^\mathrm{Pin}\right|}}{\left|\boldsymbol{\psi}_n-\boldsymbol{\psi}_l^\mathrm{Pin}\right|}e^{-j\theta_l}\sqrt{p_l}\!\right)\!s+w_n\\
&\triangleq h_n(\mathcal{S})s+w_n,
\label{received}
\end{align}
where $h_n(\mathcal{S})$ is the normalized channel of user $n$ for antenna activation. In this letter, by assuming that the transmit power of all antennas is equally allocated, and taking the power loss in the dielectric waveguide into consideration, the transmit power at the antenna activated at position $l$ is given by
\begin{equation}
p_l=\frac{P_t}{|\mathcal{S}|}10^{-\frac{\kappa \left|\boldsymbol{\psi}_0^\mathrm{Pin}-\boldsymbol{\psi}_l^\mathrm{Pin}\right|}{10}},
\end{equation}
where $\kappa$ is the average attenuation factor along the dielectric waveguide in dB/m \cite{yeh2008essence}. We note that $\kappa=0$ represents the special case of a lossless dielectric and a perfectly conducting surface.

Eq. \eqr{received} can be viewed as a multi-user downlink NOMA system, where successive interference cancelation (SIC) can be utilized to remove multiple-access interference. To this end, an SIC decoding order needs to be established based on the effective channel gains, as follows:
\begin{equation}
|h^{(1)}(\mathcal{S})|^2\le|h^{(2)}(\mathcal{S})|^2\le\cdots\le|h^{(N)}(\mathcal{S})|^2,
\end{equation}
where $|h^{(m)}(\mathcal{S})|^2$ indicates the channel condition of the $m$-th user. That is, for user $(m)$, it can first decode and remove the transmitted symbols intended for user $(i)$, $i<m$, and then decode its own symbols by treating the symbols for user $(j)$, $j>m$, as interference. Without loss of generality, assuming that the channel gain of user $n$ is the $m$-th ordered user, the achievable data rate of user $n$ is given by
\begin{equation}
R_n(\mathcal{S})=R^{(m)}(\mathcal{S})=\log_2\!\left(\!1\!+\!\frac{\alpha^{(m)}|h^{(m)}(\mathcal{S})|^2}{\sum_{i=m+1}^N\!{\alpha^{(i)}}|h^{(m)}(\mathcal{S})|^2\!+\!\sigma^2}\!\right),
\end{equation}
where $\alpha^{(m)}$ is the power allocation coefficient of the $m$-th ordered user, and $\sigma^2$ is the noise power. For the user with the strongest channel, it can successfully remove the symbols intended for the other users and decode its own symbol at the following rate:
\begin{equation}
R^{(N)}(\mathcal{S})=\log_2\!\left(\!1\!+\!\frac{\alpha^{(N)}|h^{(N)}(\mathcal{S})|^2}{\sigma^2}\!\right).
\end{equation}
\vspace{-2mm}
\subsection{Antenna Activation Problem Formulation}
The users' achievable data rates are significantly affected by the positions of the pinching antennas. That is, there exists an optimal set of activated antennas' positions $\mathcal{S}^*$, which maximizes the spectral efficiency. In order to improve the performance of the considered system, a sum rate maximization problem is formulated based on antenna activation, as follows:
\begin{subequations}
\begin{empheq}{align}
\max_{\mathcal{S}}\quad & \sum_{n\in\mathcal{N}}R_n(\mathcal{S})\\
\textrm{s.t.} \quad & \mathcal{S} \subseteq \mathcal{L},\\
& \!|\mathcal{S}|\le K.
\end{empheq}
\label{problem}
\end{subequations}\vspace{-2mm}\\
Specifically, constraints (\ref{problem}b) and (\ref{problem}c) indicate that the pinching antennas should be activated from the pre-configured positions and the number of activated antennas cannot greater than the number of available antennas, respectively.
\vspace{-2mm}
\section{Matching based Antenna Activation}
Problem \eqr{problem} is a challenging integer programming problem, and hence, in this section, matching theory is utilized to solve it. By constructing a matching between antennas $\mathcal{K}$ and positions $\mathcal{L}$, i.e., which antenna is activated at which position, the solution $\mathcal{S}^*$ can be obtained.
\vspace{-2mm}
\subsection{Formulation of One-Sided Matching}
It can be observed that in the formulated antenna activation problem, by treating the pinching antennas and the positions as two disjoint sets of players, only the pinching antennas have preferences over the positions, which is consistent with the concept of one-sided matching, i.e., housing market matching models \cite{shapley1974cores}. On the other hand, in the considered matching, both the pinching antennas and positions can remain in an unmatched state. As a result, problem \eqr{problem} can be defined as a one-sided one-to-one matching $(\mathcal{K}, \mathcal{L}, \succ, \Phi_0)$ with unmatched players, where $\succ$ is a list of strict preferences of the antennas over the positions, and $\Phi_0$ is an initial endowment matching. The corresponding definition is presented as follows:
\begin{definition}
An one-sided one-to-one matching $\Phi$ is a mapping function $\Phi: \mathcal{K}\to \mathcal{L}$ such that
\begin{enumerate}[label=(\theenumi)]
\item $\Phi(k)\in \mathcal{L}\cup\{\emptyset\}, \forall k\in\mathcal{K}$; 
\item $|\Phi(k)|=\{0,1\}, \forall k\in\mathcal{K}$; and
\item $\Phi(k)=l\Rightarrow \Phi(k')\neq l, \forall k, k' \in\mathcal{K}$,
\end{enumerate}
where $|\Phi(k)|$ denotes the cardinality of $\Phi(k)$, and $\Phi(k)=l$ denotes that antenna $k$ is paired with position $l$.
\end{definition}
In the above definition, condition (1) indicates that any pinching antenna needs to be assigned to a position or remain unmatched. Condition (2) indicates each pinching antenna can be assigned to maximum one position. Condition (3) indicates that each position cannot be occupied by more than one pinching antennas. 
\vspace{-2mm}
\subsection{Design of Matching Based Antenna Activation Algorithm}
Since the adjustment of one pinching antenna will simultaneously change the channel conditions and received signals of all users, their data rates will be affected accordingly, which means that the considered matching has externalities. That is, the preference of any player is also influenced by the preferences of the other players. In this case, the preferences of all players are constructed based on the objective function of problem \eqr{problem}, i.e., the sum rate, as follows:
\begin{equation}
U(\Phi)=\sum_{n\in\mathcal{N}}R_n(\Phi),
\end{equation}
where $R_n(\Phi)$ is user $n$'s data rate for matching $\Phi$.

Based on the defined utility, the preference list of each antenna can be established. To this end, a 2-tuple $\langle\Phi(k),\Phi\rangle$ is introduced to denote that antenna $k$ is matched with $\Phi(k)$ in matching $\Phi$. In the case that pinching antenna $k\in\mathcal{K}$ prefers position $l\in\mathcal{L}$ over its current state, this strict preference can be characterized as follows:
\begin{equation}\label{pref1}
\langle\Phi(k),\Phi\rangle \prec_k \langle l, \Phi'\rangle\ \Leftrightarrow U(\Phi) < U(\Phi').
\end{equation}
This preference means that antenna $k$ prefers to be activated at position $l$ because the utility can be strictly increased, and the matching is transformed from $\Phi$ to $\Phi'$ due to this action. In \eqr{pref1}, antenna $k$ can be deactivated in matching $\Phi$, i.e., $\Phi(k)=\emptyset$, or activated at another position, i.e., $\Phi(k)=l', l'\neq l$. In the case that pinching antenna $k$ is willing to be unmatched, the following preference is obtained:
\begin{equation}
\langle\Phi(k),\Phi\rangle \prec_k \langle\emptyset, \Phi'\rangle \Leftrightarrow U(\Phi) < U(\Phi').
\end{equation}
The above preference implies that the utility can be increased if antenna $k$ becomes deactivated, and hence, $\Phi(k)\neq \emptyset$. We note that in the considered system, swapping the matching of any two antennas (including deactivated antennas) has no impact on the utility, and therefore, the swap action is not considered in this paper.

With the defined preferences, the solution of problem \eqr{problem} can be obtained if the matching is in the core \cite{abdulkadirouglu2013matching}, which is defined as follows:
\begin{definition}\label{core}
A matching $\Phi$ is in the core if and only if there does not exist a matching $\Phi'$ such that
\begin{enumerate}[label=(\theenumi)]
\item $\langle\Phi'(k), \Phi'\rangle\succeq_k \langle\Phi(k), \Phi\rangle, \forall k \in\mathcal{K}$,
\item $\langle\Phi'(k), \Phi'\rangle\succ_k \langle\Phi(k), \Phi\rangle, \exists k \in\mathcal{K}$.
\end{enumerate}
\end{definition}

\begin{algorithm}[t]
\caption{Matching based Antenna Activation Algorithm}
\label{algorithm}
\begin{algorithmic}[1]
\STATE \textbf{Initialization:}
\STATE Randomly match pinching antennas and positions to obtain initial matching $\Phi_0$, and set $\Phi=\Phi_0$.
\STATE \textbf{Main Loop:}
\FOR{$k\in\mathcal{K}$}
\FOR{$l\in\mathcal{L}$}
\IF{$\Phi(k')\neq l, \forall k'\in\mathcal{K}$}
\IF{$\langle\Phi(k), \Phi\rangle \prec_k \langle l, \Phi'\rangle$}
\STATE Set $\Phi=\Phi'$.
\ENDIF
\ELSE
\IF{$\Phi(k)= l$}
\IF{$\langle\Phi(k), \Phi\rangle \prec_k \langle\emptyset, \Phi'\rangle$}
\STATE Set $\Phi=\Phi'$.
\ENDIF
\ENDIF
\ENDIF
\ENDFOR
\ENDFOR
\end{algorithmic}
\end{algorithm}

Based on the definitions of the preferences and the core, a matching based antenna activation algorithm is provided in Algorithm \ref{algorithm}. In this algorithm, the initial matching $\Phi_0$ is obtained by assigning the pinching antennas to random positions, where all antennas are activated with the initial matching. For the remaining steps of the algorithm, an antenna is selected in the order from $1$ to $K$ and attempts to be assigned to a position in the order from $1$ to $L$. Specifically, if target position $l$ is not occupied by other antennas, antenna $k$ attempts to be activated at this position, as shown in lines 6-9 of the algorithm. If target position $l$ is currently occupied by antenna $k$, this antenna attempts to be deactivated, as shown in lines 11-14 of the algorithm. The main loop is repeated until no antenna can be adjusted during a complete cycle. At this point, the solution of problem \eqr{problem}, i.e., $\boldsymbol{\psi}^\mathrm{Pin}$, can be obtained through the final matching.
\vspace{-2mm}
\subsection{Properties Analysis for the Proposed Algorithm}
In this subsection, the properties of the proposed matching based antenna activation algorithm are analyzed.

\subsubsection{Complexity}
The computational complexity of the proposed algorithm can be analyzed by considering the worst case, in which all pinching antennas attempt to be activated at all positions. For $K$ antennas and $L$ positions, $KL$ computations are performed in a complete cycle. For a given number of cycles $C$, the computational complexity is given by $\mathcal{O}(CKL)$, where $\mathcal{O}$ denotes the standard big-O notation.

\subsubsection{Convergence}
During the execution of the algorithm, a new matching is recorded if any antenna has been successfully assigned to a position. From initial matching $\Phi_0$, the matching is transformed by the algorithm as follows:
\begin{equation}\label{transformation}
\Phi_0 \rightarrow \Phi_1 \rightarrow \Phi_2 \rightarrow \cdots \rightarrow \Phi_\mathrm{final}.
\end{equation}
Assume that $\Phi_a$ and $\Phi_b$ are two adjacent matchings in the above sequence, where $b=a+1$. From $\Phi_a$ to $\Phi_b$, one of the pinching antennas is activated at a different position, or is deactivated. Based on the definition of the preferences, the utility is strictly increased from $\Phi_a$ to $\Phi_b$, i.e., $U(\Phi_a)<U(\Phi_b)$. Therefore, the following inequality can be obtained:
\begin{equation}
U(\Phi_0) < U(\Phi_1) < U(\Phi_2) < \cdots < U(\Phi_\mathrm{final}).
\end{equation}
Due to the fact that the number of antennas and locations are finite, the number of possible matchings is limited. Given the strict preference, the proposed matching based algorithm is guaranteed to converge to a final matching whose utility cannot be further increased.

\subsubsection{Stability}
Assuming that the final matching $\Phi_\mathrm{final}$ in \eqr{transformation} is not a stable matching, based on Definition \ref{core}, there exist at least one matching in which an antenna can be assigned to a different position or deactivated. However, this case contradicts the fact that $\Phi_\mathrm{final}$ is the final matching and its utility cannot be further increased. Therefore, the final matching obtained by Algorithm \ref{algorithm} is always stable.
\vspace{-2mm}
\section{Simulation Results}
In this section, the performance of the considered NOMA assisted pinching-antenna system employing the proposed matching based antenna activation algorithm is investigated. In the simulations, the parameters are set as follows: $d=3$~meters, $f_c=28$~GHz, $n_\mathrm{eff}=1.4$, and $\sigma^2=-90$~dBm. Moreover, a benchmark based on fixed power allocation is considered, where the available transmit power of each activated antenna is equally distributed to all signals, i.e., $\alpha_n=1/N, \forall n\in\mathcal{S}$.

\begin{figure}[!t]
\centering{\includegraphics[width=80mm]{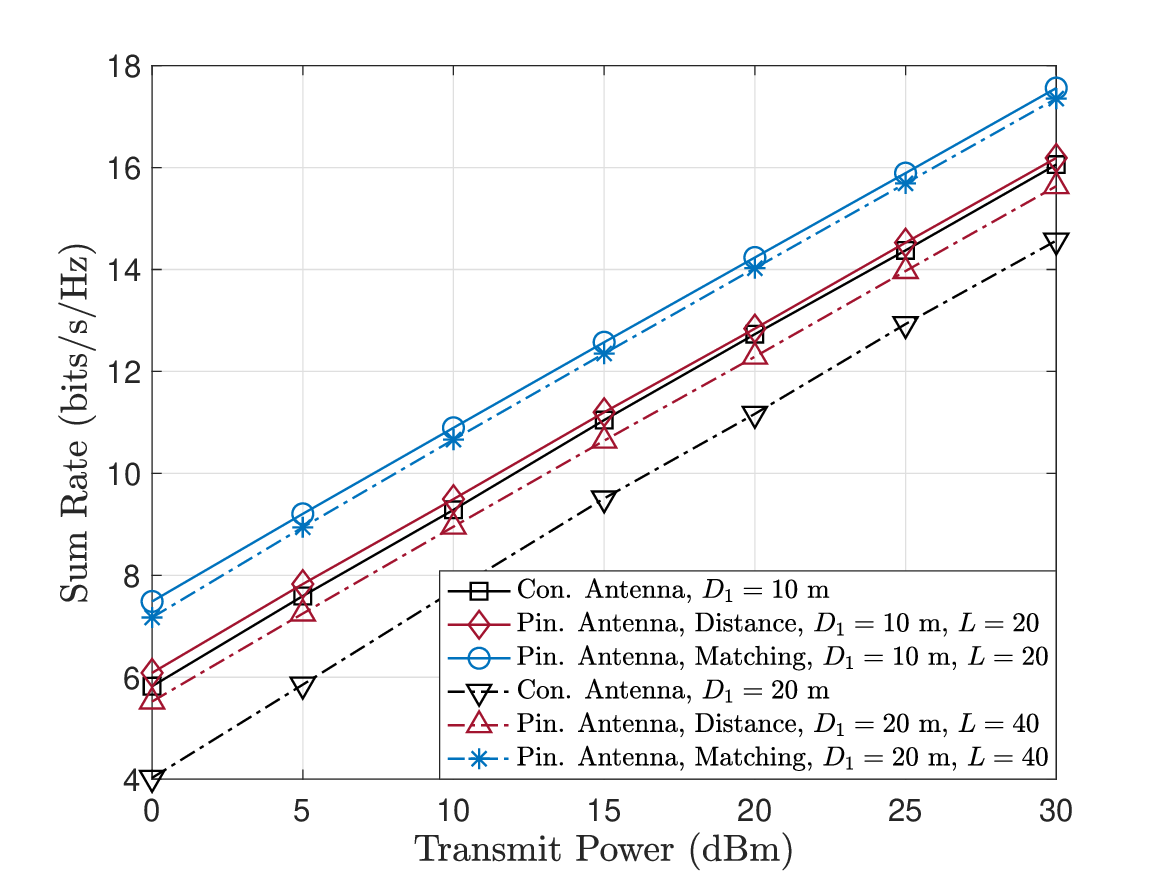}}
\caption{Impact of the transmit power on the sum rate, where $D_2=4$~m, $K=4$, $N=4$, and $\kappa=0.1$~dB/m.}
\label{result1}
\vspace{-4mm}
\end{figure}

\begin{figure}[!t]
\centering{
\subfigure[Sum Rate]{\centering{\includegraphics[width=80mm]{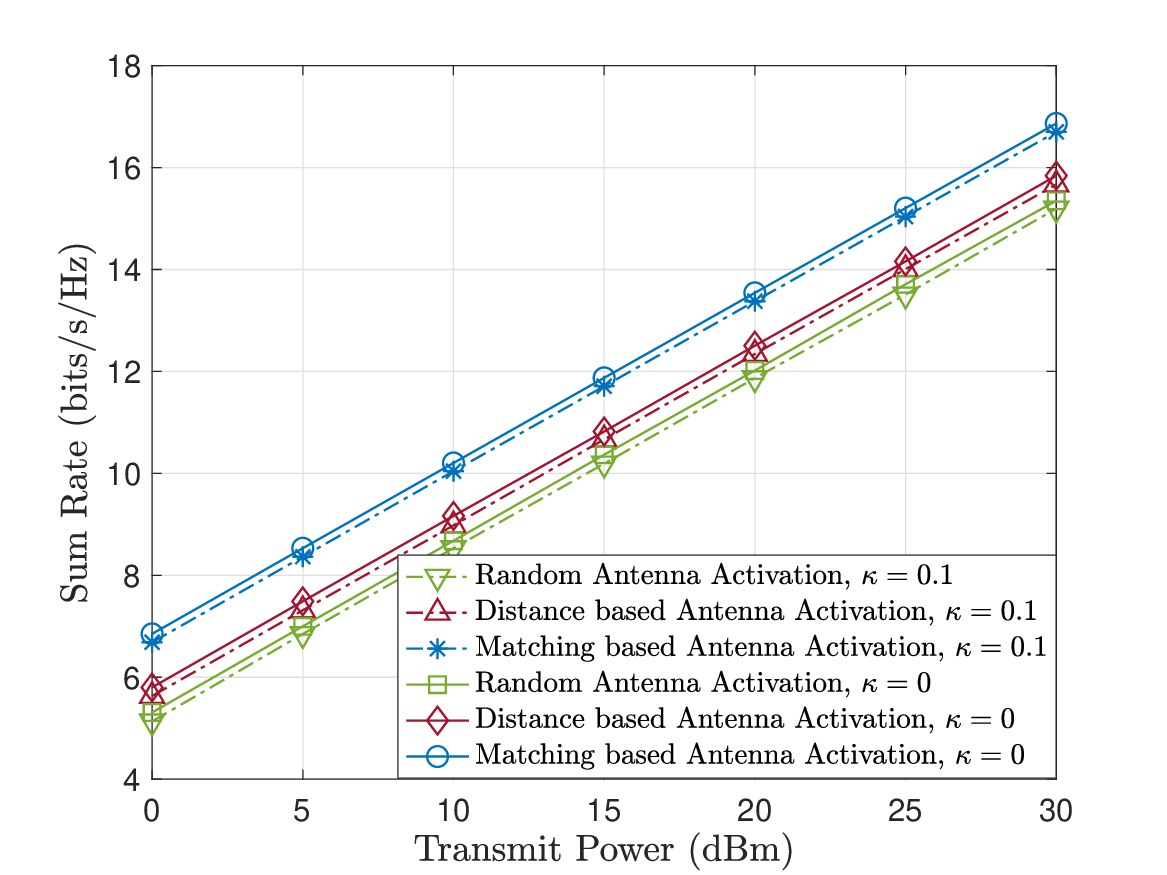}}}\vspace{-1.5mm}
\subfigure[Fairness]{\centering{\includegraphics[width=80mm]{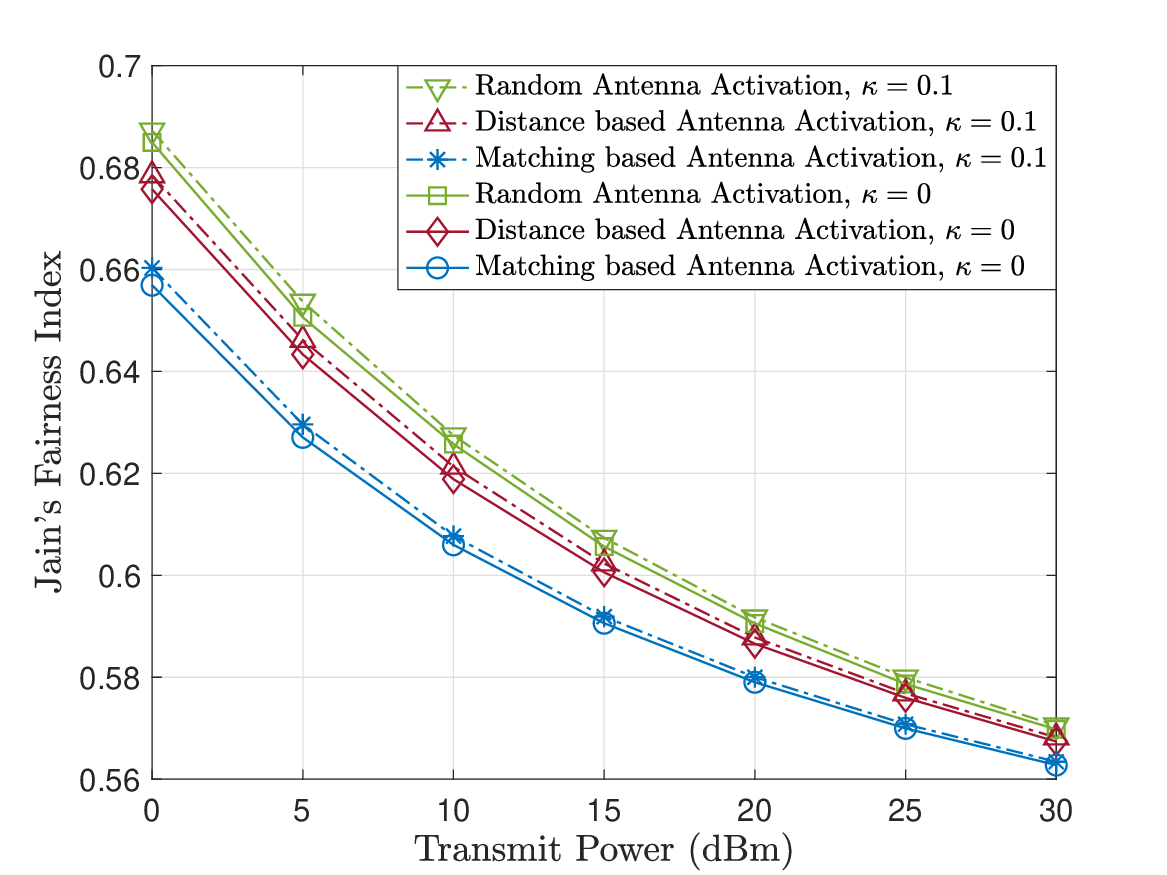}}}}\vspace{-2mm}
\caption{Achievable sum rate in the considered pinching-antenna system with different antenna activation schemes, where $D_1=10$~m, $D_2=6$~m, $K=2$, $N=2$, and $L=20$.}
\label{result2}
\vspace{-4mm}
\end{figure}

\fref{result1} shows the performance of the pinching-antenna system and compares it to that of a conventional antenna system, in which the $K$ antennas are located in the centre with height $d$ and antenna space $\lambda/2$. Moreover, distance based antenna activation shown in \cite{ding2024pin} is used as a benchmark, where each pinching antenna is activated at the location closest to the corresponding user, i.e., $\psi_k^\mathrm{Pin}=(x_n,0,d), \forall k=n$. It can be observed that the considered pinching-antenna system outperforms the conventional antenna system in terms of sum rate, and the gap increases significantly with the length of the rectangular area. This is because in conventional systems, an increase in length leads to the deterioration of the channel gains, while in pinching-antenna systems only the propagation loss in the waveguide is increased, which is insignificant compared to free-space path loss. Moreover, it is worth noting that matching based antenna activation can dynamically adjust the locations of the activated antennas according to the propagation loss, which ensures that the performance of the pinching-antenna system is less affected.

In \fref{result2}(a), the sum rate gain achieved by the proposed algorithm is studied, where random antenna activation is included as a benchmarking scheme, in which the initial matching $\Phi_0$ is directly used. As can be observed, the sum rate increases as the transmit power grows for all schemes, and the proposed matching based antenna activation algorithm achieves the best performance. Recall that the strategy of this algorithm is to improve the effective channel gain of the strongest user who can contribute the most to the sum rate. This strategy is confirmed by \fref{result2}(b), where matching based antenna activation has the lowest fairness. Furthermore, in the case that the propagation loss in the waveguide is non-zero, the sum rate of all schemes is only slightly reduced, which indicates that the dominant factor in pinching-antenna systems is the free-space path loss.

\begin{figure}[!t]
\centering{\includegraphics[width=80mm]{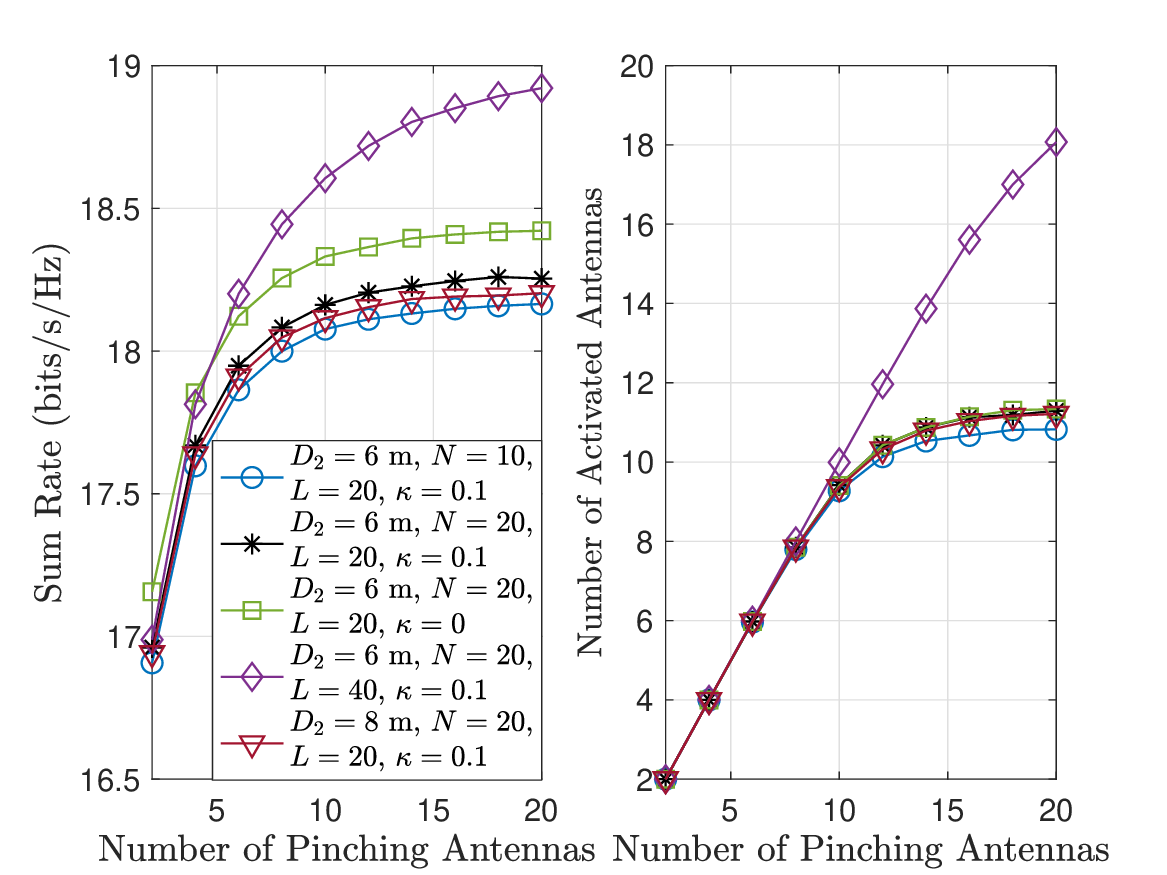}}
\caption{Impact of the number of antennas in the considered pinching-antenna system with the proposed matching based antenna activation, where $D_1=10$~m, and $P_t=30$~dBm.}
\label{result3}
\vspace{-4mm}
\end{figure}

\fref{result3} illustrates the dependence of the sum rate on the number of activated antennas for different scenarios. Considering the case with $D_2=6$~m, $N=L=20$, and $\kappa =0.1$~dB/m as a baseline,  it can be seen from the figure that as the number of users decreases, both the sum rate and the number of the required pinching antennas decrease. When the propagation loss is neglected (or the short side $D_2$ is increased), the sum rate increases (or decreases), while the effect on the number of activated antennas is insignificant. On the other hand, if the number of available positions grows, the sum rate is further increased and more antennas are activated, since the pinching antennas can be adjusted more precisely.

\begin{figure}[!t]
\centering{\includegraphics[width=80mm]{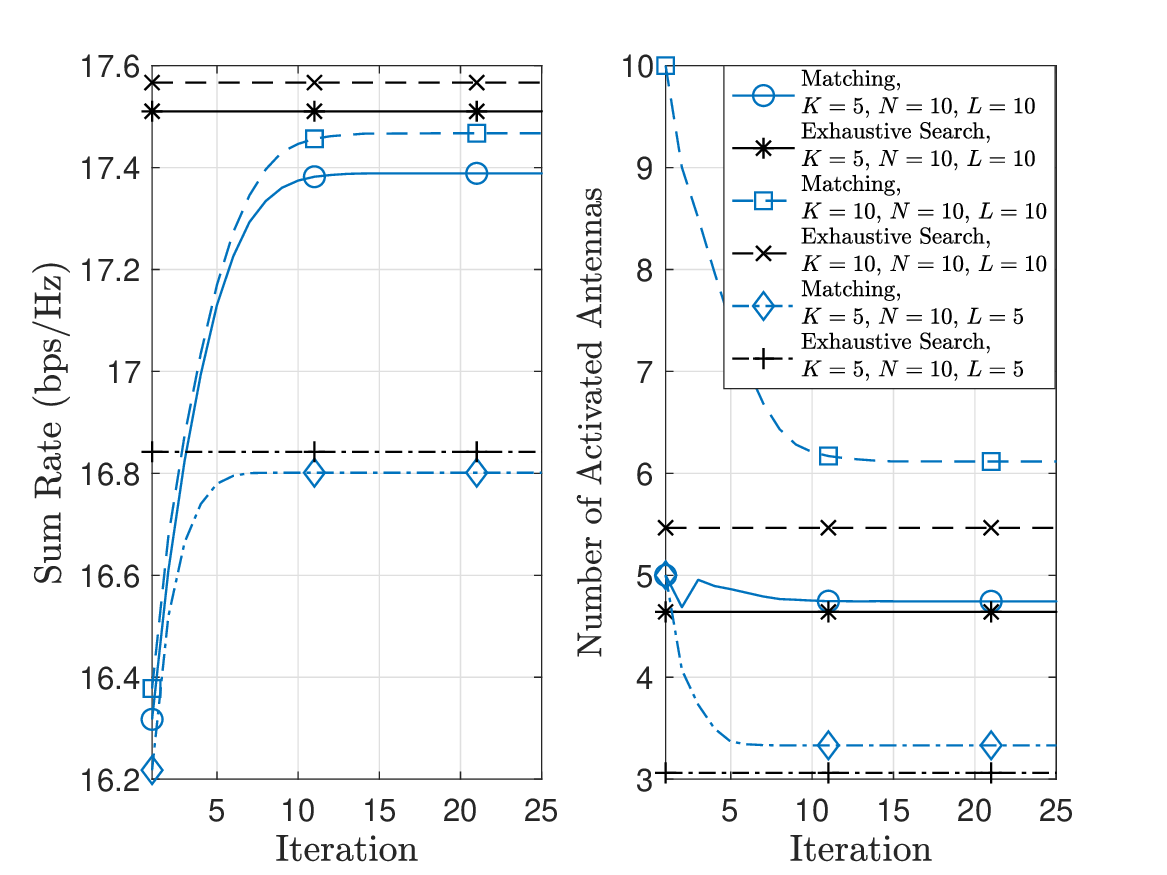}}
\caption{Convergence performance of the proposed algorithm, where $D_1=10$~m, $D_2=6$~m, $\kappa=0.1$~m/dB, and $P_t=30$~dBm.}
\label{reult4}
\vspace{-4mm}
\end{figure}

The convergence performance of the proposed matching based antenna activation algorithm is shown in \fref{reult4}, where an exhaustive search is included as baseline. It can be observed that the matching based algorithm can achieve around $99\%$ of the performance of the global optimal solution within $20$ iterations. Therefore, the proposed algorithm can be regarded as a low-complexity near-optimal solution. According to the results of the exhaustive search, it can be found that a higher sum rate can be obtained with less pinching antennas, which shows the importance of optimizing the number of activated antennas. Furthermore, based on \fref{reult4}, the analysis of the convergence and stability presented at the end of the previous section can be verified.
\section{Conclusions}
In this letter, antenna activation was studied in a NOMA assisted pinching-antenna system. By introducing several potential positions for the pinching antennas, a sum rate maximization problem was formulated and treated as a one-sided matching problem. Based on this, a low-complexity matching based antenna activation algorithm was proposed to obtain a near-optimal solution. Furthermore, the benefits of the considered pinching-antenna system and the proposed matching based algorithm were unveiled via the presented simulation results.
\bibliographystyle{IEEEtran}
\bibliography{KaidisBib}
\end{document}